\documentclass[11pt]{article}

\usepackage{amsmath}
\usepackage[psamsfonts]{amssymb}
\usepackage{amsthm}
\usepackage{xspace}

\newcommand{\for}{\quad \text{for }}
\newcommand{\Span}{\mathrm{Span}}
\newcommand{\End}{\mathrm{End}\xspace}
\newcommand{\Hom}{\mathrm{Hom}\xspace}
\newcommand{\tr}{\mathrm{Tr}}
\newcommand{\Div}{\mathrm{div}\,}

\newcommand{\C}{{\mathbb C}}
\newcommand{\Z}{{\mathbb Z}}

\newcommand{\N}{{\mathbb N}}
\newcommand{\PP}{{\mathbb P}}
\newcommand{\HH}{{\mathcal H}}
\newcommand{\OO}{{\mathcal O}}
\newcommand{\A}{{\mathcal A}}
\newcommand{\I}{{\mathcal I}}

\newcommand{\nn}{{\mathbf n}}
\newcommand{\uu}{{\mathbf u}}

\newcommand{\abs}[1]{\lvert#1\rvert}

\newcommand{\IP}[1]{\langle#1\rangle}
\newcommand{\iso}{\simeq}
\newcommand{\done}{$\hfill \hfill \blacksquare$ \bigskip}
\newcommand{\dwrt}[1]{\frac{\partial}{\partial#1}}

\newcommand{\ci}{\oint}

\newtheorem{defn}{Definition}
\newtheorem{thm}[defn]{Theorem}
\newtheorem{lem}[defn]{Lemma}
\newtheorem{cor}[defn]{Corollary}
\newtheorem{prop}[defn]{Proposition}

\newcommand{\oproof}[1]{\noindent {\bf Proof#1.\ }}

\newcommand{\ti}[1]{\textit{#1}}

\newcommand{\ie}{{\it i.e.}}
\newcommand{\eg}{{\it e.g.}}

\begin{document}

\def\thefootnote{\fnsymbol{footnote}}
\setcounter{page}{1}
\pagestyle{plain}

\begin{flushright}
  hep-th/0009235 \\
  DAMTP-2000-111
\end{flushright}

\begin{center} \LARGE{\textbf{Rationality, quasirationality 
and finite W-algebras}} \end{center}
\vskip 1.0cm
\begin{center}{\large Matthias R.\ Gaberdiel}%
\footnote{{\tt e-mail: mrg@mth.kcl.ac.uk}}%
\footnote{Address after 1 October 2000: Department of Mathematics,
King's College London, Strand, London WC2R 2LS, U.K.}  
and {\large Andrew Neitzke}%
\footnote{{\tt e-mail: aneitzke@alumni.princeton.edu}}%
\footnote{Address after 1 October 2000: Harvard University, Department
of Mathematics, One Oxford Street, Cambridge, MA 02138, USA.} \\ 
Department of Applied Mathematics and Theoretical Physics \\
Cambridge University \\
Wilberforce Road \\
Cambridge CB3 0WA\\
United Kingdom\\\end{center}

\begin{center} September 2000 \end{center}

\vskip 0.8cm

\begin{abstract}
Some of the consequences that follow from the $C_2$ condition of Zhu
are analysed. In particular it is shown that every conformal field
theory satisfying the $C_2$ condition has only finitely many $n$-point
functions, and this result is used to prove a version of a conjecture
of Nahm, namely that every representation of such a conformal field
theory is quasirational. We also show that every such vertex operator
algebra is a finite $W$-algebra, and we give a direct proof of the
convergence of its characters as well as the finiteness of the fusion
rules. 
\end{abstract}

\section{Introduction} \label{introsec}

Conformal field theory has had a major impact on modern theoretical
physics as well as modern mathematics. From the point of view of 
physics, conformal field theory plays a central r\^ole in 
string theory, at present the most promising candidate for a unifying
theory of all forces. On the other hand, conformal field theory
inspired the purely mathematical definition of vertex operator algebras,
which has led to beautiful and deep results in the 
theory of finite groups and number theory.

The significance and power of conformal field theory was first
conclusively demonstrated by the work of Belavin, Polyakov and
Zamolodchikov \cite{BPZ}. They fixed a general framework for its study  
which was further developed by Moore and Seiberg \cite{MS2}, in
particular. On the other hand, the mathematical theory of vertex
operator algebras is due to Borcherds \cite{Borch,Borch2} and Frenkel,
Lepowsky \& Meurman \cite{FLM}, and was further developed by Frenkel, 
Huang \& Lepowsky \cite{FHL}, Zhu \cite{Zhu}, Kac \cite{Kac} and
others. Apart from this algebraic viewpoint, there exists also a
geometrical approach that was directly inspired by string theory (in
particular the work of Friedan \& Shenker \cite{FS}) and that has been
put on a mathematical foundation by Segal \cite{Segal} and Huang
\cite{Huang1,Huang2,Huang3}. 

Much has been learned about conformal field theory, but there are
still a number of conceptual problems that have not been
resolved so far. One of them concerns the question of how to
characterise the class of `rational' theories, \ie\ those theories
that are in some sense finite and tractable. Various definitions of 
rationality have been proposed, but the interrelations
between the different assumptions are not very well understood.

One of the assumptions that were introduced by Zhu in \cite{Zhu} in
order to be able to prove the convergence of the characters is the
condition (sometimes referred to as the $C_2$ condition) that a
certain quotient space of the vertex operator algebra is
finite-dimensional. This is a slightly technical assumption; however,
it has the great virtue of being easily testable in concrete
examples. In this paper we analyse the consequences that follow from
this condition. As we shall show, the $C_2$ condition implies that a
whole family of quotient spaces are finite-dimensional, and this in
turn is sufficient to prove that the theory has only finitely many
$n$-point functions (Theorem~\ref{findim}), and in particular that all
fusion rules between irreducible highest weight representations are
finite (Corollary~\ref{fusion}).  We also prove that the $C_2$
condition implies that each highest weight representation of the
theory is quasirational (Theorem~\ref{quasirat}); this proves a
version of a conjecture of Nahm \cite{Nahm}. Finally, we show that
every such vertex operator algebra is a finite $W$-algebra, and we
give a direct proof of the convergence of its characters (see also
\cite{DLM2, Li} for independent proofs of these results.) Most of
these results hinge on finding a small spanning set of the vacuum
representation (Proposition~\ref{core}); we expect that this result
may also be useful in other contexts and applications.

If we assume in addition that Zhu's algebra is semisimple, we can also
find an upper bound on the (effective) central charge of the theory in
terms of the dimension of the $C_2$ quotient space
(Proposition~\ref{cbo}). 
\smallskip

The paper is organised as follows. In Section~\ref{notsec} we fix our
conventions and introduce some notation. In Section~\ref{Ann} we
define a class of quotient spaces that generalise the construction of
the $C_2$ space (and of Zhu's algebra), and we prove a number of
simple properties. In Section~\ref{reps} we recall the definition of a
highest weight representation and we explain in which sense Zhu's
algebra classifies these representations. Section~\ref{ratl} is
concerned with the different definitions of rationality. The central
Proposition is proven in Section~\ref{basislem}, and a few simple
consequences are derived. In Section~\ref{nconj} we use this result to
prove that each conformal field theory satisfying the $C_2$ condition
has only finitely many $n$-point functions, and we show that this
implies Nahm's conjecture. Section~\ref{centralcharge} describes our
bound on the central charge. Section~\ref{interp} gives a more precise
description of one of the quotient spaces which figure prominently in
the proofs of the preceding results, and Section~\ref{conclus} contains some
conclusions and outlook for further work. Finally, we have included an
appendix where two quite technical calculations are described in
detail.

\section{Notation} \label{notsec}

We assume the reader is familiar with basic notions of conformal field
theory, as found for instance in \cite{DFMS,Greview}. Some
acquaintance with the language of vertex operator algebras
\cite{Borch,FLM,Kac} is also helpful.  

At all times in this paper we are considering a fixed chiral bosonic
conformal field theory defined on the sphere $\PP$. To be
precise, by a ``chiral bosonic conformal field theory'' we
mean an object of the type discussed in \cite{GG}. From this point of
view, a conformal field theory on $\PP$ is defined in terms of its
amplitudes. We assume that the amplitudes are local, M\"obius
invariant, and satisfy the cluster decomposition property (that
guarantees that the spectrum of the scaling operator $L_0$ is bounded
by zero from below, with a unique state $\Omega$ of eigenvalue
$h=0$). We also assume that the theory is conformal, \ie\ that it
possesses a stress energy tensor $V(L,z)=L(z)$ whose modes $L_n$
satisfy the Virasoro algebra. The details of the chosen formalism are
not essential to following the ideas of this paper, and indeed, the
whole argument could be rewritten in terms of the standard axioms of
vertex operator algebras \cite{Kac}.

The amplitudes that define the theory are written as
$\IP{\prod_{a=1}^k V(\psi_a, z_a)}$, where the vertex operator
corresponding to $\psi\in V$ is denoted by $V(\psi, z)$. The vector
space $V$ consists of quasi-primary states that generate the whole 
theory; for convenience we shall 
occasionally assume that $V$ {\it is} the space of all quasiprimary
states. We sometimes write $V(\psi,z)$ in terms of modes as 
\begin{equation}
V(\psi, z) = \sum_{n \in \Z} V_n(\psi) z^{-n-h_\psi} = \sum_{m\in\Z} 
V_{(m)} (\psi) z^{-m-1} \,,
\end{equation}
where $V_{(m)}(\psi)=V_{m+1-h_\psi}(\psi)$ is the moding that is
commonly used in the mathematical literature. 

We will frequently consider meromorphic functions and differentials
defined on the Riemann sphere $\PP$.  It is convenient to use the
language of ``divisors'' (see \cite{GH}) to classify the zeros and
poles of these functions.  We now give a brief review of the facts
relevant for us. A divisor on $\PP$ is, by definition, any formal sum
of the form 
\begin{equation}
D = \sum_{P \in \PP} c_P [P]\,, \qquad c_P \in \Z\,, \qquad
\textrm{finitely many } c_P \ne 0\,. 
\end{equation}
Divisors can be added and subtracted in the obvious way, and we say 
$D\ge 0$ if all $c_P \ge 0$. Now let $\nu_P(f)$ denote the order of
vanishing of $f$ at $P$ (so $\nu_P(f)$ is negative if $f$ has a pole
at $P$), and define the divisor of $f$ to be 
\begin{equation}
\Div f = \sum_{P \in \PP} \nu_P(f) [P]\,.
\end{equation}
Clearly $\Div fg = \Div f + \Div g$, and $\Div f \ge 0$ just if $f$ is
holomorphic (\ie\ constant).

We can similarly define $\Div \omega$ where $\omega$ is a meromorphic
$k$-differential on $\PP$; explicitly, such an $\omega$ can always be
written as $\omega = f dz^{\otimes k}$ for some $f$, and then we have
\begin{equation}
\Div \omega = \Div f + k \Div dz = \Div f - 2k [\infty]\,.
\end{equation}
(This definition expresses the fact that $dz$ has a pole of order $2$
at infinity.) 

The crucial analytic property which the amplitudes of the theory must
possess by definition \cite{GG} is that, for any $\psi \in V$, 
$\IP{V(\psi, z) \prod_{i=1}^{k} V(\psi_i,z_i)} dz^{\otimes h_\psi}$
depends meromorphically on $z \in \PP$
and has poles only for $z=z_i$.

The Fock space $\HH$ of the theory is spanned by finite linear
combinations of states of the form 
\begin{equation}\label{Fock}
\Psi= V_{(n_1)}(\psi_1) \cdots V_{(n_k)}(\psi_k) \Omega \,,
\end{equation}
where $\psi_i\in V$, $\Omega$ denotes the unique (vacuum) state with
conformal weight $h=0$, and $n_i\in\Z$.  Any product of vertex
operators $V(\phi_1, u_1)\cdots V(\phi_l,u_l)$ defines a linear
functional on the Fock space by
\begin{equation}\label{functional}
\eta_{V(\phi_1, u_1)\cdots V(\phi_l,u_l)}(\Psi) = 
\ci_0 dz_1 z_1^{n_1} \cdots \ci_0 dz_k z_k^{n_k}
\IP{V(\phi_1, u_1)\cdots V(\phi_l,u_l) \prod_{i=1}^{k}
V(\psi_i,z_i)}\,,
\end{equation}
where the contours are chosen so that $|z_1|>|z_2|>\cdots >|z_k|$.
The Fock space is the space spanned by vectors of the form
(\ref{Fock}), modulo states that vanish in each linear functional
associated to any product of vertex operators. In (\ref{functional})
we have considered the Fock space at $0 \in \PP$; however, since the
amplitudes are translation invariant, it is clear that one can
similarly consider the Fock space at any other point on the Riemann
sphere.

\section{The subspaces $A_\nn$} \label{Ann}

We begin by introducing a generalization of the quotients of $\HH$
which appeared in \cite{GG,N,Zhu}. For fixed 
$\uu = (u_1, \dots, u_k)\in (\PP - \{0\})^k$ 
(where we do not require that the $u_i$ be distinct), we define
\begin{equation} \label{defO}
\begin{split}
O_\uu = \Span \ \Big\{ \ci_0 dz \, g(z) V(\psi, z) \chi \ \Big\vert\ 
       &\chi \in \HH, \psi \in V,\, \hbox{$g$ meromorphic}, \\ 
       & \Div g dz^{\otimes -h_\psi+1} \ge -N[0] + \sum_{i=1}^k h_\psi
       [u_i] \textrm{    for some } N\geq 0\Big\}\,. 
\end{split}
\end{equation}
We then set 
\begin{equation} \label{defA}
A_\uu = \HH / O_\uu\,.
\end{equation}
Because of the M\"obius invariance of the amplitudes, we may assume
that one of the $u_i$, $u_1$ say, is equal to $\infty$. If none of the
other $u_j$ are equal to $\infty$, one can give an explicit
description of $O_\uu$ as the space spanned by the states of the form
$V_\uu^{(M)}(\psi)\chi$ with $M>0$, where  
\begin{equation}\label{Npoint}
V_\uu^{(M)}(\psi) = \oint_0 \frac{d\zeta}{\zeta^{M+1}} V\left[
\left(\frac{\prod_{j=2}^{k}(\zeta-u_j)}{\zeta^{k-2}}\right)^{L_0} 
\psi,\zeta\right] 
\end{equation}
and $\psi \in V$, $\chi\in\HH$. For the case where the $u_i$ are
distinct, this space has been considered before in \cite{GG}, where it
was denoted by $A_{k}$ (but we now renounce that notation in favour
of one described below.)
In the case where all $k$ of the $u_i$ equal
$\infty$, one can give a similarly explicit description: in this case
$O_\uu$ is simply spanned by states of the form
$V_{-N-(k-1)h_\psi}(\psi)\chi$ with $N>0$.  This choice of $\uu$ is
particularly convenient since the resulting $O_\uu$ is spanned by
states of definite conformal weight; this makes calculations
significantly simpler. 

The original motivation for the definition of $A_\uu$, in the case
where all $u_i$ are distinct, stemmed from the fact that the algebraic
dual space $A_\uu^*$ describes the correlation functions involving $k$
highest weight states at $u_1,\dots, u_k$ (this was first observed by
Zhu in \cite{Zhu} for $\uu = (-1, \infty)$). More generally, one finds

\begin{thm} \label{zthm} There is a one-to-one linear correspondence
between elements $\eta \in A_\uu^*$ and systems of correlation
functions on the sphere, \ie\ maps
\begin{equation}
((\psi_1, z_1), \dots, (\psi_l, z_l)), \psi_i \in V, z_i \in \PP \quad
\mapsto \quad \IP{\prod_{j=1}^l V(\psi_j, z_j)}_\eta \in \C
\end{equation}
such that the $\IP{\prod_{j=1}^l V(\psi_j, z_j)}_\eta$ (regarded only
as functions of the $z_j$) obey the operator product relations of the
theory defined on the sphere (see \cite{GG} for a precise definition),
and have the ``highest weight'' property
\begin{equation} \label{analprop}
\Div \IP{V(\psi,z)}_\eta \, dz^{\otimes h_{\psi}} \ge 
- \sum_{i=1}^k h_{\psi} [u_i]\,. 
\end{equation}
\end{thm}
\oproof{} Given any system of correlation functions one can construct
a linear functional $\eta$ on $\HH$ by contour integration, as
discussed in Section \ref{notsec}. If we further require
\eqref{analprop}, then this functional vanishes on $O_\uu \subset\HH$,
and therefore $\eta \in A_\uu^*$. Conversely, any $\eta \in A_\uu^*$ 
defines formal Laurent series, whose convergence to functions with the
required analytic properties was proven in \cite{N}. (Strictly
speaking the proof was only given under the additional hypothesis that
the $u_i$ be distinct; however, that hypothesis is actually not
required anywhere in the proof.)
\done 

It is often convenient to use a shorthand notation where we only keep
track of the number of coincident points $u_i$. Let us thus
define $A_\nn$ where $\nn$ is a multi-index $\nn=[n_1,\dots,n_l]$;
this denotes the space $A_\uu$ for the case where $n_1$ of the $u_i$
are equal to $v_1$, $n_2$ of the $u_i$ are equal to $v_2\ne v_1$, 
{\it etc.}  We define $X_{\nn}$ to be the corresponding configuration
space, namely the set of all $\uu \in \PP^{\abs{\nn}}$ (where 
$\abs{\nn}=n_1+n_2+\cdots+n_l$) for which the first $n_1$ coordinates 
are coincident, the second $n_2$ coordinates are coincident, and so
on. The usefulness of this notation depends on the following fact:   

\begin{thm} \label{ptindep} Suppose $A_{(\infty^k)}$ is
finite-dimensional and let $\abs{\nn}=k$.  Then the space $A_\nn$ is
independent of the choice of $\uu \in X_{\nn}$, in the sense that
choosing a homotopy class of paths from $\uu$ to $\uu'$ in $X_{\nn}$
determines a natural isomorphism $A_\uu \iso A_{\uu'}$.
\end{thm}
\oproof{} Using Theorem~\ref{zthm} we can regard $A_\nn^*$ as a space
of correlation functions. First consider the case 
$\nn = (1, 1, \dots,1)$ where all $u_i$ are distinct. In that case we
can introduce a more suggestive notation for the correlation
functions, namely, we write 
\begin{equation}\label{suggestive}
\IP{\prod_{i=1}^k W(\phi_i, u_i) 
\prod_{i=1}^{k} V(\psi_i,z_i)} \equiv 
\IP{\prod_{i=1}^{k} V(\psi_i,z_i)}_\eta\,. 
\end{equation}
(Here the formal symbols $W(\phi_i, u_i)$ represent insertions of
highest weight states.) Given all correlation functions at some fixed
$\uu$, the  Knizhnik-Zamolodchikov equation determines them at all
$\uu$ using the fact that the Virasoro algebra acts geometrically;
more specifically, if $u_1 \ne \infty$ then 
\begin{equation} \label{kz}
\dwrt{u_1} \IP{\prod_{i=1}^k W(\phi_i, u_i)} 
= \ci_{u_1} dz \IP{\prod_{i=1}^k W(\phi_i, u_i) L(z)}\,.
\end{equation}
Using this formula systematically one can construct a family of
differential equations, to be solved in the space obtained by gluing
together the $A_\uu^*$ at different points $\uu$; if these equations
admit solutions, we then expect that they will define the analytic
continuation from correlation functions at $\uu$ to correlation
functions at $\uu'$, proving the theorem. 

To prove that solutions actually exist one has to impose the condition
that $A_{(\infty^k)}$ is finite-dimensional (specifically, what one
uses is the fact that a basis for $A_{(\infty^k)}$ corresponds to a
spanning set for each $A_\uu$, $\uu \in X_\nn$.)  Under this
assumption it is shown in \cite{N} that the $A_\uu^*$ (and hence the
$A_\uu$) indeed fit together to form a vector bundle over $X_\nn$
which possesses a natural flat connection given by \eqref{kz}. The
argument given there extends straightforwardly to the case where the
$u_i$ need not be distinct. \done  

The space $O_{(\infty, \infty)}$ is the $C_2$ space of Zhu, so if
$A_{[2]}$ is finite-dimensional, the $C_2$ condition of Zhu is
satisfied. On the other hand, $A_{[1,1]}$ is isomorphic to Zhu's
algebra (compare also Section~\ref{reps}). 

The space $A_{[k]}$ has been considered before in \cite{N}, where it
was denoted by $\HH/C_{k}$.  As we now show, its dimension provides an
upper bound on the dimension of the spaces $A_\nn$ with $\abs{\nn}=k$;
this result was already used in \cite{Zhu} for the special case 
$\nn= (1,1)$ (see also \cite{N}).   

\begin{lem} \label{ptsplit} 
$\dim A_\nn \le \dim A_{[\abs{\nn}]}$.   
\end{lem}
\oproof{} Fix $\uu=(\infty,u_2,\dots,u_k)$ (by M\"obius invariance
this involves no loss of generality).  It follows from (\ref{defO})
that $O_\uu$ is generated by the states of the form  
\begin{equation}
V_{\uu}^{(N)}(\psi) \chi =
\sum_{s=0}^{(k-1)h_\psi} c_s V_{-N-(k-1)h_\psi+s}(\psi) \chi 
\end{equation}
where $N>0$ and $c_s$ are some constants (depending on $\uu$) with 
$c_0=1$. On the other hand, $O_{(\infty^k)}$ is generated by the
states of the form $V_{-N-(k-1)h_\psi}(\psi)\chi$ with $N>0$. 

Let $\{\phi_1,\dots,\phi_M\}$ be a set of representatives for $\HH$
modulo $O_{(\infty^k)}$.  We claim
that these vectors also span $\HH$ modulo $O_\uu$.  Suppose that this
is not the case, and let $\Psi$ be a vector of minimal conformal
weight that does not differ by an element in $O_\uu$ from a linear
combination of $\phi_1,\dots,\phi_M$.  By assumption we can write 
\begin{equation}
\Psi = \sum_{j=1}^{M} b_j \phi_j 
+ \sum_{r=1}^{L} V_{-N_r-(k-1)h_r}(\psi_r)\chi_r\,.
\end{equation}
But then 
\begin{equation}
\widehat{\Psi} = \Psi -  \sum_{j=1}^{M} b_j \phi_j  - 
\sum_{r=1}^{L} V_{\uu}^{(N_r)}(\psi_r)\chi_r
\end{equation}
is a linear combination of vectors whose conformal weight is strictly
smaller than that of $\Psi$. By the minimality of $\Psi$ it then
follows that $\widehat{\Psi}$ differs by an element in $O_\uu$ from a
linear combination of $\phi_1,\dots,\phi_M$, and we have the desired
contradiction. 
\done

It should be noted that the dimension can actually decrease when we
`split points'. The simplest example for this phenomenon occurs
already for $\nn = [1,1]$: the $e_8$ level $1$ theory is self-dual 
(\ie\ the only representation is the vacuum representation), and
therefore has no nontrivial two-point functions, implying 
$\dim A_{[1,1]}=1$; on the other hand, it is easy to see
by inspection that $\dim A_{[2]} \geq 249$.

\section{Representations and Zhu's algebra} \label{reps}

We now shift from considering the vacuum representation $\HH$ to more
general representations of the conformal field theory.

A representation of the conformal field theory is defined in terms
of the amplitudes it induces \cite{GG,Zhu},
\begin{equation}\label{repamp}
\langle W(\phi_1,u_1) W(\phi_2,u_2) \prod_{i=1}^{k}
V(\psi_i,z_i)\rangle \,,
\end{equation}
where the $\psi_i\in V$ are arbitrary. The amplitudes have the crucial
property that they respect the operator product relations of the
meromorphic conformal field theory. Furthermore, the amplitudes are
M\"obius covariant, and are analytic as a function of the $z_i$,
except for possible poles at $z_i=z_j$, $i\ne j$, and singularities at
$z_i=u_j$. We call the representation amplitudes \ti{non-singular} if 
the singularities at $z_i=u_j$ are poles of finite order; a
non-singular representation amplitude is \ti{highest weight} if the 
order of the pole at $z_i=u_j$ is bounded by $h_{\psi_i}$. 

We can construct from the amplitudes two vector spaces $\HH^1$ and
$\HH^2$ which form modules for the modes $V_n(\psi)$ for $\psi\in\HH$. 
These modules are generated by the action of the modes (defined via
contour integrals around the $u_i$) from $\phi_1$ and $\phi_2$,
respectively. The actual module is then a quotient space of the space
so obtained, where we remove ``null vectors'' by identifying states
whose difference vanishes in all amplitudes \eqref{repamp}.  

The requirement that the amplitudes respect the operator product
relations implies that the action of the modes satisfies the ``Jacobi
identity'' required in algebraic definitions of representation (given
\eg\ in \cite{Kac}). If the representation amplitudes are
non-singular, then the two representations are ``weak modules'' in the
sense of \cite{DLM}, \ie\ $\HH^i$ has the property that, for any
$\psi\in \HH$ and $\chi \in\HH^i$, $V_n(\psi)\chi = 0$ for $n>N$
(where $N$ may depend on $\psi$, $\chi$). Finally, if the
representation amplitudes are highest weight, then the two
representations are highest weight representations, \ie\ $\HH^i$ is
generated from a single state $\phi_i$ with the property that
$V_n(\psi)\phi_i = 0$ whenever $n>0$. 

On the other hand, we can construct representation amplitudes (that
have the appropriate analytic properties) from purely algebraic
data. Indeed, it follows from Theorem~\ref{zthm} that each element in
the algebraic dual of $A_{(u_1,u_2)}$ defines representation
amplitudes that have the highest weight property. It was furthermore
shown by Zhu \cite{Zhu} (see also \cite{Greview,GG} for an exposition
more in line  with the present point of view) that $A_{(u_1,u_2)}$ has
the structure of an algebra, and that the equivalence classes of
representation amplitudes (where we identify amplitudes that define
equivalent modules $\HH^i$) are in one-to-one correspondence with
representations of $A_{(u_1,u_2)}$.  Finally, the irreducible
representations $R$ of Zhu's algebra are in one-to-one correspondence
with the irreducible highest weight representations $\HH^R$ of the
conformal field theory.

The algebra structure of Zhu's algebra is most easily understood for
$\A=A_{(\infty, -1)}$, whence it is defined by 
\begin{equation}
\psi * \chi \equiv V_{(\infty,-1)}^{(0)}(\psi) \chi\,,
\end{equation}
where $V_{(\infty,-1)}^{(0)}(\psi)$ is given in (\ref{Npoint}).
This product is characterized by the identity (emphasized by Brungs
and Nahm in \cite{NB})  
\begin{equation}
V_0(\psi * \chi) = V_0(\psi) V_0(\chi)
\end{equation}
which holds when both sides act on highest weight states, so that $\A$
is essentially the algebra of zero modes of fields in the vacuum
sector acting on highest weight states.
\smallskip

Theorem~\ref{zthm} states that $A^*_{(u_1, \dots, u_k)}$ describes the
space of correlation functions that correspond to $k$ highest weight 
states. If we are, however, interested in understanding the different
ways in which the various representations of the theory can couple in
$k$-point functions, then this description contains a certain
redundancy. In particular, we can act with zero modes $V_0(\psi)$ on
any of the $\phi_i$ in \eqref{suggestive}, and this will produce
another highest weight state in the same representation. It is
therefore useful to study $A^*_\uu$ as a representation of $k$ copies
of the zero mode algebra $\A$ acting at the $k$ points $u_i$. 

\begin{thm} \label{aacts} Fix a multi-index $\nn$.  For any $i$ with
$n_i=1$ there is a natural map of algebras, $\rho_i:\A\to\End(A_\nn)$.
The dual map $\rho_i^*: \A \to \End(A_\nn^*)$ satisfies the identity 
(for $u_i\ne\infty$) 
\begin{equation}\label{ident}
\IP{\cdots}_{\rho_i^*(\psi) \eta} = \ci_{u_i} dz (z-u_i)^{h_\psi - 1}
\IP{V(\psi, z)\cdots}_\eta\,, 
\end{equation}
\ie\ it is the action of zero modes.
\end{thm}

\oproof{} Without loss of generality we may assume that
$\uu=(\infty,u_2,\dots,u_k)$. Using the notation introduced in
(\ref{Npoint}) we define $\rho_1(\psi)=V_{\uu}^{(0)}(\psi)$. It is
shown in the appendix that for $L>0$
\begin{equation}\label{eq1}
[V_\uu^{(0)}(\psi_1), V_\uu^{(L)}(\psi_2)] \chi \in  O_\uu \,,
\end{equation}
and that for $L\geq 0$,
\begin{equation}\label{eq2}
V_{\uu}^{(0)}\left(V_{(\infty,-1)}^{(L)}(\psi)\,\chi\right)\,\phi 
\approx V_{\uu}^{(L)}(\psi) \,V_\uu^{(0)}(\chi) \,\phi\,,
\end{equation}
where $\approx$ denotes equality up to states in $O_\uu$. This implies
that $\rho_1$ defines an algebra homomorphism $\A \to \End(A_\nn)$. 
Using the M\"obius invariance of the amplitudes, this is sufficient to
prove the statement for all $i$.  The formula \eqref{ident} follows
easily from the definition of the action. \done

If $\A$ is semisimple and if all $u_i$ are distinct,
Theorem~\ref{aacts} allows us to decompose $A^*_{(u_1,\dots, u_k)}$
completely into representations $(R_1 \otimes \cdots\otimes R_k)$ of
$\A^k$; the multiplicity with which $(R_1 \otimes \cdots\otimes R_k)$
appears in $A^*_{(u_1,\dots,u_k)}$ then gives an upper bound on the
number of different ways in which the spaces 
$\HH^{R_1}, \dots,\HH^{R_k}$ can be coupled.\footnote{For theories in
which every representation is completely reducible (see
Section \ref{ratl}) this bound is sharp, \ie\ every element of
$A^*_\uu$ corresponds to an actual coupling. The reason this is not
true in general is that the correlation functions coming from an
element of $A^*_\uu$ need not respect the null-vector relations in the
$\HH^{R_i}$.} 
\medskip

Given a representation, a rough measure of its size relative to the
vacuum representation is given by the \ti{special subspace}, defined 
by Nahm in \cite{Nahm} as follows: let $W \subset \HH^i$ be defined
by 
\begin{equation}
W = \Span\{V_n(\psi) \chi: n \le -h_\psi<0, \psi \in \HH,
\chi \in \HH^i\}\,.
\end{equation}
Then a special subspace, $\HH^i_s$, is a subspace of $\HH^i$ such
that $W + \HH^i_s = \HH^i$ and $W \cap \HH^i_s = \{0\}$. The dimension
of $\HH^i_s$ equals the dimension of the quotient space $\HH^i/W$, and
thus is independent of the choice of $\HH^i_s$. In the case of 
the vacuum representation, $\dim \HH_s=1$, and $\dim \HH^i_s >1$ for
any other representation. Representations whose special subspace is
finite-dimensional play a preferred role (their fusion rules are
finite), and are called \ti{quasirational}.   

Finally, since $\HH^i$ carries an action of the $V_n(\psi)$, we note
that we can define various quotients $A_\nn^i$ of $\HH^i$ just by
replacing $\HH$ with $\HH^i$ in \eqref{defO}, \eqref{defA}.  In
particular, when $\HH^i = \HH^R$, $A_{[1]}^R$ is isomorphic to the
highest weight space $R$, as can be seen from choosing $\uu =(\infty)$
in \eqref{defO}.  The $A_\nn^R$ obey an analogue of Theorem
\ref{zthm}, but we will not use this fact explicitly in what follows.

\section{Rationality} \label{ratl}

One of the central concepts in conformal field theory is
`rationality,' a condition which is supposed to express a kind of
finiteness of the theory. There exist various notions of finiteness in
the literature \cite{DLM,Li,MS1,Zhu} and the precise interrelations
between the different assumptions are not all understood. On the other
hand, most people would agree that every rational theory should have
the following properties:
\begin{list}{(\roman{enumi})}{\usecounter{enumi}}
\item The conformal field theory has only finitely many irreducible 
highest weight representations. 
\item The characters $\chi_R(q) = \tr_{\HH^R} q^{L_0}$ are
convergent for $\abs{q} < 1$ and close under modular transformations.
\item The fusion rule coefficients $N_{ij}^k$ of three irreducible
highest weight representations, $\HH^i$, $\HH^j$ and $\HH^k$ are all
finite. 
\end{list}

There are various different conditions that imply some of these
properties. For example, if Zhu's algebra is semisimple, it 
follows from the Wedderburn structure theorem (see for example
\cite{FD}) that 
\begin{equation} \label{wedd}
\A = \bigoplus_{i} \End V_i
\end{equation}
for a finite set of finite-dimensional vector spaces $V_i$, which form
the only irreducible representations of $\A$. Thus if $\A$ is
semisimple, (i) is satisfied. It is reasonable to conjecture that (ii)
and (iii) should also follow from the semisimplicity of $\A$, but this
conjecture is, at least at present, still out of reach. 

\noindent In order to make progress, two other conditions have been
proposed:
\begin{list}{(\alph{enumi})}{\usecounter{enumi}}
\item Every $\N$-graded weak module is completely reducible. (This is
the condition called rationality by Zhu and many other authors on
vertex operator algebras \cite{DLM, Li, Zhu}.)
\item The quotient space $A_{[2]}$ is finite-dimensional. (This
is the $C_2$ condition of Zhu.)
\end{list}
It has been shown in \cite{Zhu} that (a) implies the semisimplicity of
$\A$, and therefore by the above argument (i).  In the same paper it
was shown that (a) together with (b) imply (ii).  Zhu further
conjectured that (a) implies (b), but this also seems at present out
of reach. The $C_2$ condition implies that $\A$ is finite-dimensional,
but does not imply its semisimplicity \cite{GK}.

In the following we shall mainly analyse the implications of (b).  In
particular we shall show that (b) implies that every highest weight
representation is quasirational and that (iii) holds. We shall also
give a direct argument for the convergence of the characters under the
assumption of (b).

\section{The basis lemma} \label{basislem}

First we will prove three computational results that are
originally due to Borcherds \cite{Borch} (see also \cite{Kac}).

\begin{lem} \label{lem1} We have 
\begin{equation}
\label{lemma1}
[V_{(-N_1)}(\psi_1), V_{(-N_2)}(\psi_2)] = \sum_{r=1}^{h_1 + h_2} 
V_{(-N_1-N_2+1-r)}(\chi_r) \,,
\end{equation}
where $h_i$ is the conformal weight of $\psi_i$, and the conformal
weight of $\chi_r$ is $h_1+h_2-r$.
\end{lem}

\oproof{} The commutator 
\begin{equation}
[V_{-N_1+1-h_1}(\psi_1),V_{-N_2+1-h_2}(\psi_2)] = \sum_{s=0}^{h_1+h_2-1}
V_{-N_1-N_2+2-h_1-h_2} (\chi_s)\,,
\end{equation}
where the conformal weight of $\chi_s$ is $h_1+h_2-1-s$. Substituting
$r=s+1$, we then obtain the above formula. \done

\begin{lem} \label{lem2} We have
\begin{equation}
\begin{split}
\label{lemma2}
V_{(-N_1)}\left(V_{(-N_2)}(\psi)\chi\right) & =
\sum_{L\geq 0} \binom{N_2+L-1}{L} 
V_{(-N_2-L)}(\psi) V_{(-N_1+L)}(\chi) \\
& \qquad 
+ (-1)^{N_2+1} \sum_{L\geq 0} \binom{N_2+L-1}{L} 
V_{(-N_1-N_2-L)}(\chi) V_{(L)}(\psi) 
\end{split}
\end{equation}
where both sums terminate when they are evaluated on an element of 
$\HH$. 
\end{lem}

\oproof{} We rewrite $V_{(-N_1)}(V_{(-N_2)}(\psi) \chi)$ as
\begin{equation}
\begin{split}
V_{(-N_1)}(V_{(-N_2)}(\psi)\chi)&=
\oint_0 V\left(V_{(-N_2)}(\psi)\chi,\zeta\right) \zeta^{-N_1} d\zeta \\
&=\oint\oint_{|\zeta|>|z|}
V\left(V(\psi,z)\chi,\zeta\right) 
z^{-N_2} \zeta^{-N_1} dz d\zeta \\
&=\oint\oint_{|\zeta|>|z|}
V(\psi,z+\zeta)V(\chi,\zeta)
z^{-N_2} \zeta^{-N_1} dz d\zeta\,.
\end{split}
\end{equation}
We then substitute $\omega=z+\zeta$ and find
\begin{equation}
\begin{split}
V_{(-N_1)}(V_{(-N_2)}(\psi)\chi)
&= {\displaystyle \oint_0\left\{\oint_\zeta
V(\psi,\omega) V(\chi,\zeta)
(\omega-\zeta)^{-N_2} d\omega\right\} 
\zeta^{-N_1} d\zeta} \\
&={\displaystyle \oint\oint_{|\omega|>|\zeta|} V(\psi,\omega)
\left\{ V(\chi,\zeta) 
{(\omega-\zeta)^{-N_2} \zeta^{-N_1}}d\zeta \right\}
d\omega}\\
&\qquad {\displaystyle - \oint\oint_{|\zeta|>|\omega|} V(\chi,\zeta)
\left\{ V(\psi,\omega) 
{(\omega-\zeta)^{-N_2}} d\omega\right\} 
\zeta^{-N_1} d\zeta\,.}
\end{split}
\end{equation}
In the first line we can then write
$$
(\omega - \zeta)^{-N_2} = \omega^{-N_2} 
\sum_{L=0}^{\infty} \binom{N_2+L-1}{L}
\left(\frac{\zeta}{\omega} \right)^L \,,
$$
and thus obtain
\begin{equation}
\begin{split}
& = \sum\limits_{L=0}^\infty \binom{N_2+L-1}{L}
\oint_0 V(\psi,\omega) \omega^{-N_2-L} d\omega
\oint_0 V(\chi,\zeta) \zeta^{-N_1+L} d\zeta \\
& = \sum\limits_{L=0}^\infty \binom{N_2+L-1}{L}
V_{(-N_2-L)}(\psi) V_{(-N_1+L)}(\chi) \,.
\end{split}
\end{equation}
Finally, we rewrite the second line as 
$$
(\omega - \zeta)^{-N_2} = (-1)^{N_2} 
\zeta^{-N_2} \sum_{L=0}^{\infty} \binom{N_2+L-1}{L}
\left( \frac{\omega}{\zeta} \right)^L \,, 
$$
and obtain
\begin{equation}
\begin{split}
& = (-1)^{N_2+1} \sum\limits_{L=0}^\infty 
\binom{N_2+L-1}{L} \oint_0 V(\chi,\zeta)
\zeta^{-N_1-N_2-L} d\zeta
\oint_0 V(\psi,\omega) \omega^{L} d\omega \\
& = (-1)^{N_2+1} \sum_{L=0}^\infty \binom{N_2+L-1}{L}
V_{(-N_1-N_2-L)}(\chi) V_{(L)}(\psi) \,.
\end{split}
\end{equation}
This proves the claim. \done

\begin{lem} \label{lem3} As an immediate corollary of
Lemma~\ref{lem2}, we have 
\begin{equation}
\begin{split}
V_{(-N)}(\psi) V_{(-N)}(\chi) & = V_{(-2N+1)}(V_{(-1)}(\psi) \chi)
- \sum_{L\geq 0, L\ne N} V_{(-1-L)}(\psi) V_{(-2N+1+L)}(\chi) \\
& \qquad - \sum_{M\geq 0} V_{(-2N-M)}(\chi) V_{(M)}(\psi)\,,
\end{split}
\end{equation}
where again both sums terminate when they are evaluated on an element
of $\HH$.
\end{lem}

\oproof{} This follows from Lemma~\ref{lem2} with $N_1=2N-1$ and
$N_2=1$. \done

The next proposition is the core of this section.  Recall that
$A_{[2]} \iso \HH/ O_{(\infty, \infty)}$ and that
$O_{(\infty,\infty)}$ is spanned by states of the form 
$V_{(-M)}(\rho)\chi$ where $\rho,\chi \in \HH$ and $M > 1$.

\begin{prop} \label{core} Let $\{W_i\}$ be a
set of representatives for $\HH$ modulo $O_{(\infty, \infty)}$. Then
$\HH$ is spanned by the set of states
\begin{equation} \label{spanset}
V_{(-N_1)} (W_{i_1}) \cdots V_{(-N_n)}(W_{i_n}) \Omega\,,
\end{equation}
where $N_1 > N_2 > \cdots > N_n > 0$.
\end{prop}

\oproof{} Define a filtration on $\HH$,
\begin{equation}
\HH^{(0)} \subset \HH^{(1)} \subset \cdots \subset \HH^{(g)} \subset
\cdots \subset \HH, 
\end{equation}
as follows:  $\HH^{(g)}$ is the subspace spanned by all states of the
form 
\begin{equation} \label{monom}
V_{(-N_1)} (\psi_1) \cdots V_{(-N_n)}(\psi_n) \Omega
\end{equation}
where $\sum_i h_{\psi_i} \le g$.  Clearly $\HH = \cup_g \HH^{(g)}$
(since every $\Psi$ has at least the trivial representation 
$\Psi = V_{(-1)} (\Psi) \Omega$, so that if $\Psi$ is homogeneous we
have $\Psi \in \HH^{(h_\Psi)}$.)

Two properties of this filtration will be useful in what follows.
First, commutator terms always have lower grade: more precisely, let
$\Psi \in \HH$ be some state of the form \eqref{monom}, with $\sum_i
h_{\psi_i} \le g$, and let $\Psi_R$ be the state obtained from $\Psi$
by exchanging two adjacent modes in \eqref{monom}.  Then 
$\Psi -\Psi_R \in \HH^{(g-1)}$, as follows readily from
Lemma~\ref{lem1}. Second, elements of $O_{(\infty, \infty)}$ decrease
the grade: again let $\Psi \in \HH$ be of the form \eqref{monom}, with
$\sum_i h_{\psi_i} \le g$, but this time with the additional
stipulation that some $\psi_i \in O_{(\infty, \infty)}$, \ie\
$\psi_i= V_{(-M)} (\rho) \chi$, $M > 1$.  Then using Lemma~\ref{lem2}
we find that $\Psi \in \HH^{(g-1)}$, since the state
$V_{(-M)}(\rho)\chi$ is of weight $h_\chi + h_\rho + (M - 1)$. 
\smallskip

\noindent For any pair $(g, N)$ of nonnegative integers we now
consider the proposition: 

\smallskip
\noindent \textbf{Inductive hypothesis.}  \textit{The space
$\HH^{(g)}$ is 
spanned by states of the form 
\begin{equation} \label{ihyp} 
V_{(-N_1)} (W_{i_1}) \cdots V_{(-N_n)} (W_{i_n}) \Omega
\end{equation}
where $N_1\ge N_2\ge \cdots\ge N_n > 0$, $\sum_j h_{W_{i_j}} \le g$, 
and $N_i = N_{i+1}$ is allowed only for $N_i > N$.} 
\smallskip

We consider pairs to be ordered lexicographically:
so $(g, N) < (g', N')$ if either $g < g'$, or $g = g'$ and $N < N'$.
Then the set of pairs is well ordered (every non-empty subset has a
smallest member).  So we can proceed by induction:  fixing $(g, N)$ we
assume the hypothesis holds for all smaller pairs and establish it
for $(g, N)$. 

In particular, the inductive hypothesis means the proposition is true
for $(g-1, N)$ so that every $\Psi \in \HH^{(g-1)}$ can be expressed
in the claimed form (this is true even for $g=0$ since in that case
$\HH^{(g-1)} = 0$.) As remarked above, provided we begin with
monomials \eqref{monom} with $\sum h_{\psi_i} \le g$, commutator terms
and terms involving states in $O_{(\infty, \infty)}$ will always be in
$\HH^{(g-1)}$; so in trying to reduce some state \eqref{monom} with
$\sum h_{\psi_i}\le g$ to the claimed form we are always free to reorder
modes and to replace any $V_{(M)}(\psi)$ by $V_{(M)}(W)$ (here and
below, we suppress the index on $W_i$, which plays no role.)

We consider separately the pairs $(g,N)$ with $N=0$. In this case,
given an element of $\HH^{(g)}$ of the form \eqref{monom}, we can put
it in the claimed form simply by reordering modes into descending
order and replacing all $\psi_i$ by $W$.  (If any mode $V_{(M)}(W)$
with $M \ge 0$ appears, it will annihilate the vacuum after the
reordering.)  

Now suppose $N>0$ and consider $\Psi$ of the form \eqref{monom} with
$\sum h_{\psi_i}\le g$. Using the inductive hypothesis applied to $(g,
N-1)$ we can write $\Psi$ as a sum of states of the form
\begin{equation} \label{bigform}
V_{(-M_1)}(W) \cdots V_{(-M_m)}(W) [V_{(-N)}(W)]^s 
V_{(-L_1)}(W) \cdots V_{(-L_l)}(W) \Omega \,,
\end{equation}
where $M_1 \ge \cdots \ge M_m > N > L_1 > \cdots > L_l \ge 0$, 
$s \ge 0$. If $s < 2$ then \eqref{bigform} is already a state of the
desired sort. If $m \ne 0$ then the expression 
$[V_{(-N)}(W)]^s\cdots \Omega$ is in $\HH^{(g-1)}$ and we can use the
inductive hypothesis applied to $(g-1, N)$ to replace it, obtaining a 
sum of expressions which have no repeated indices at or below $N$.  On
the other hand, if $m=0$ and $s \ge 2$  then we use Lemma~\ref{lem3}
to replace the initial pair $V_{(-N)}(W) V_{(-N)}(W)$. This
replacement generates two sorts of terms:  first, it generates 
\begin{equation} \label{semibadterm}
V_{(-2N+1)}(\psi) [V_{(-N)}(W)]^{s-2} 
V_{(-L_1)}(W) \cdots V_{(-L_l)}(W) \Omega\,, 
\end{equation}
second, it generates
\begin{equation} \label{badterm}
V_{(- N - K)}(\psi) V_{(- N + K)}(\chi) [V_{(-N)}(W)]^{s-2}
V_{(-L_1)}(W) \cdots V_{(-L_l)}(W) \Omega \,,
\end{equation}
where $K>0$ (using our freedom to reorder modes.)  As usual we are
free to replace $\psi, \chi$ by $W$ everywhere. Now omitting the first
mode from \eqref{semibadterm} or \eqref{badterm} produces a state
$\Psi' \in \HH^{(g-1)}$ (unless the first $W$ is actually the vacuum, 
which can happen in \eqref{semibadterm} in the special case $N=1$ ---
we treat this case separately below).  Using the inductive hypothesis 
for $(g-1, N)$ we then rewrite $\Psi'$ in terms of monomials
\eqref{ihyp} with no repeated indices at or below $N$.  This yields
the desired result, since $2N-1$  and $N+K$ are both greater than $N$,
so that re-attaching the omitted mode does not generate a repeat at or
below $N$.  

It only remains to consider \eqref{semibadterm} in the special case
$N=1$.  In this case we can rewrite that term simply as
$[V_{(-1)}(W)]^{s-1} \Omega$, and repeat the process until we are left
with $V_{(-1)}(W) \Omega$.  This completes the proof of the inductive
hypothesis for all $(g,N)$.

To complete the proof of the proposition we use the fact that $\HH$ is
graded by conformal weight, $\HH = \cup_{h\geq 0} \HH_h$, where
$\HH_h$ consists of states of weight
$h$. It is therefore sufficient to show that each $\HH_h$ is spanned
by states \eqref{spanset} with 
$N_1>N_2>\cdots > N_n>0$.  But this follows directly from the
inductive hypothesis together with the fact that the conformal weight
of the state in \eqref{ihyp} is greater than or equal to 
$\sum_j (N_j-1)$; thus if \eqref{ihyp} is of weight $h$,  
none of the $N_i$ can be greater than $h+1$, so the result
follows by choosing $N=h+1$ and sufficiently large $g$ in the inductive
hypothesis. This completes the proof. \done

We remark that the spanning set given by Proposition~\ref{core} is not
actually a basis; this can be seen already for the minimal model with
$c=-22/5$, for which the set $\{W_i\}$ can be taken to be 
$\{\Omega,L_{-2}\Omega\}$.  Then \eqref{spanset} includes both 
$L_{-3} L_{-2} \Omega$ and $L_{-5} \Omega$, but in fact these two 
states are linearly dependent. Nevertheless, Proposition~\ref{core} is
a very useful tool as we shall see momentarily.

Most of the known conformal field theories are generated by a finite
set of quasiprimary fields, and are indeed what is called \ti{finite
$W$-algebras}. More precisely, a vertex operator algebra is a finite
$W$-algebra if it contains a finite set of states $W_i\in\HH$,
$i=1,\dots, n$, such that $\HH$ is spanned by states of the form
\eqref{ihyp} where $N_1\geq N_2 \geq \cdots \geq N_n>0$ and
$i_j\geq i_{j+1}$ whenever $N_j = N_{j+1}$. It now follows directly from  
Proposition~\ref{core} that

\begin{cor}\label{finalg} If $A_{[2]}$ is finite-dimensional, then
the vertex operator algebra is a finite $W$-algebra.
\end{cor}
\oproof{} We take the $\{W_i\}\in\HH$ to be a set of representatives
for $\HH$ modulo $O_{(\infty, \infty)}$ and apply Proposition~\ref{core}.
\done

It is sometimes assumed in the definition of a vertex operator algebra
that each $L_0$ eigenspace is finite-dimensional. It now follows
directly from Proposition~\ref{core} that this is automatic provided
that $A_{[2]}$ is finite-dimensional.

Actually, Corollary~\ref{finalg} has been proven before in \cite{Li}.
The generating set Li used was somewhat different, however.  He
defined a space $C_1 \subset \HH$ and then showed that $\HH$ is
spanned by all states $V_{n_1}(\psi_1) \cdots V_{n_m}(\psi_m) \Omega$,
where the $\psi_i$ range over some complementary subspace to $C_1$.
This result was then refined in \cite{KL} where it was observed that
the modes can actually be taken in a fixed lexicographical order;
furthermore it was shown that $\HH / C_1$ is a ``minimal'' generating
set in a certain sense.  These results are actually stronger than
our Corollary~\ref{finalg} because finite-dimensionality of $\HH /
C_1$ is much weaker than our hypothesis.  On the other hand, our
spanning set has the significant advantage that it allows us to prove
the ``no repeat'' condition of Proposition~\ref{core}, which will be
critical in the arguments of Sections \ref{nconj} and
\ref{centralcharge}.

The next result has also been obtained before, in \cite{DLM2}:

\begin{prop}\label{findimm} If $A_{[2]}$ is finite-dimensional then
the character 
\begin{equation}\label{chardef}
\chi(q) = \tr_{\HH} q^{L_0 - \frac{c}{24}}\,,
\end{equation}
which is defined as a formal power series, converges for
$0<\abs{q}<1$. 
\end{prop}
\oproof{} Let us denote by $Q(n,k)$ the number of partitions of $n$
into integers of $k$ colours, with no integer appearing twice in the
same colour. Then Proposition~\ref{core} implies that
\begin{equation}\label{ineqq}
\tr_{\HH} \; q^{L_0} \le 
\sum_{n \ge 0} q^n Q(n,k) = 
\prod_{n > 0}  (1 + q^n)^k\,, 
\end{equation}
where the inequality holds for each coefficient of the power series
and hence for real positive $q$. (We set $k = \dim A_{[2]} - 1$ rather
than $k = \dim A_{[2]}$ because we can always choose one of the $W_i$ 
to be $\Omega$, and $V_{(-N)}(\Omega)=\delta_{N,1} \mathbf{1}$.) The
right-hand-side converges for $0<\abs{q}<1$ since the modulus of its
logarithm is bounded by 
\begin{equation}
k \sum_{n=1}^{\infty} \abs{\log (1+q^n)}
\leq k \sum_{n=1}^{\infty} \frac{\abs{q}^n}{(1-\abs{q}^n)} 
\leq  \frac{k}{(1-|q|)} \sum_{n=1}^{\infty} \abs{q}^n \,.
\end{equation}
By the comparison test this then implies the convergence of the
character $\chi(q)$ for $0<\abs{q}<1$. \done

We remark that by similar techniques to those used in the proof of
Proposition~\ref{core} one can show that $\HH^R$ is spanned 
by the states of the form (see also \cite{KL} for a similar argument)
\begin{equation} \label{spanset2}
V_{-N_1} (W_{i_1}) \cdots V_{-N_n}(W_{i_n}) U_i\,,
\end{equation}
where $U_i$ runs over a basis of the highest weight space $R$ of
$\HH^R$, and $N_1 \ge N_2 \ge\cdots \ge N_n> 0$. If the representation
in question is irreducible, $\dim A_{[2]}<\infty$ implies that $R$ is 
finite-dimensional, and we can bound the character of the
representation $\HH^R$ (defined in analogy to \eqref{chardef})
by
\begin{equation}
\chi_R(q) \le (\dim R) q^{-\frac{c}{24}}
\left(\prod_{n=1}^{\infty}(1-q^n)\right)^{-k} \,.
\end{equation}
This is again sufficient to prove the convergence of these characters
for $0<\abs{q}<1$.

\section{Nahm's conjecture} \label{nconj}

In this section we will be exploring some further consequences of the
assumption that $A_{[2]}$ is finite-dimensional.  We remark that
results similar to those appearing in this section have been proven in
\cite{Li}, under the assumption that $L_0$ acts semisimply on all weak
modules.  This assumption is somewhat difficult to check in practice,
however, and in any case is strictly stronger than
finite-dimensionality of $A_{[2]}$.\footnote{The triplet algebra
\cite{GK} satisfies the $C_2$ condition, but it possesses
representations for which $L_0$ does not act semisimply.}

We shall first prove that every conformal field theory for which
$A_{[2]}$ is finite-dimensional possesses only finitely many $n$-point
functions. Given Theorem~\ref{zthm} this statement follows from the
following observation.

\begin{thm}\label{findim} Suppose $A_{[2]}$ is finite-dimensional.
Then all $A_\uu$ are finite-dimensional. 
\end{thm}
\oproof{} By Lemma~\ref{ptsplit} we see that it is sufficient to
show that all $A_{(\infty^k)}$ are finite-dimensional.  By definition,
\begin{equation}
O_{(\infty^k)} = \Span\{V_{(-M)}(\rho) \chi: \rho \in \HH, \chi \in \HH,
M > (k-2)h_\rho+1\}\,.
\end{equation}
Now consider the spanning set for $\HH$ provided by
Proposition~\ref{core}.  Since $A_{[2]}$ is assumed finite-dimensional
we can choose the set $\{W_i\}$ to be finite.  So $\HH$ is spanned by 
monomials 
\begin{equation} \label{sstate}
V_{(-N_1)}(W_{i_1}) \cdots V_{(-N_n)}(W_{i_n}) \Omega\,,
\end{equation}
where $N_1 > \cdots > N_n > 0$.  But if 
$N_1 > (k-2)\textrm{max}\{h_{W_i}\}+1$ then the state \eqref{sstate}
is in $O_{(\infty^k)}$.  This leaves us only finitely many choices for
the $N_i$, which gives a finite spanning set for $\HH/O_{(\infty^k)}$,
completing the proof. \done

Now we are in a position to prove Nahm's conjecture. Let $\HH^R$ be
some irreducible highest weight representation of the conformal field
theory. In \cite{Nahm} Nahm defined the special subspace $\HH^R_s$ (as
discussed in Section~\ref{reps}) and defined $\HH^R$ to be
\ti{quasirational} if $\HH^R_s$ is finite-dimensional. Nahm
conjectured that the rationality of the theory implies that all
irreducible representations are quasirational. We shall now prove this
statement under the condition that $A_{[2]}$ is finite-dimensional.   

In fact, we shall prove a slightly stronger statement, namely that 
all quotient spaces $A_{[n]}^R$ are finite-dimensional. This implies
that all representations are quasirational since 
$\dim A_{[2]}^R \ge \dim \HH^R_s$, because
\begin{equation}
A_{[2]}^R \iso A^R_{(\infty, \infty)} 
= \HH^R / \Span\{V_n(\psi) \chi: n<-h_\psi,\psi\in V,\chi\in\HH^R\}\,.
\end{equation}

The motivation for our proof comes from the interpretation of the
quotients $A_\nn$ as spaces of correlation functions. From
Theorem~\ref{findim} and Lemma~\ref{ptsplit} we know that $A_{[2]}$
finite-dimensional implies $A_{[p,1]}$ finite-dimensional for all
$p\geq 1$; and from Theorem \ref{zthm} we know that $A_{[p,1]}^*$ can
be understood as the space of correlation functions $\IP{\cdots}_\eta$
with the property that 
\begin{equation}
\Div \IP{V(\psi, z)}_\eta dz^{\otimes h_\psi} \ge - p h_\psi [u_1] -
h_\psi [u_2]\,. 
\end{equation}
But this analytic structure is exactly what we would expect from
correlation functions that are induced by a single highest weight
state at $u_2$ and a state at $u_1$ that is annihilated by all
$V_{n}(\psi)$ with $n>(p-1) h_\psi$. If we choose $u_1=\infty$,
$u_2=0$, the state at $u_1=\infty$ defines a linear functional on the
Fock space $\HH^R$ at $u_2=0$. The property that the state at $u_1$ is
annihilated by the modes with $n>(p-1) h_\psi$ implies then that this
functional vanishes on $O_{(\infty^p)}^R \subset \HH^R$, and therefore
defines a functional on $A_{(\infty^p)}^R$. We therefore expect that
we can construct an element of $A_{[p,1]}^*$ from a highest weight
state $U$ in the representation $R$, and an element 
$\eta\in (A_{[p]}^R)^*$; more specifically, if we evaluate the linear
functional in $A_{[p,1]}^*$ on $\chi\in\HH$ (now regarding $\HH$ as
being placed at $1 \in \PP$) we should have   
\begin{eqnarray}
\IP{\eta(\infty) \chi(1) U(0)} & = & 
\sum_{n \in \Z} \IP{\eta(\infty) (V_{-n}(\chi) U)(0)} \\
& = & \sum_{n = 0}^{(p-1)h_\chi} \IP{\eta(\infty) (V_{-n}(\chi) U)(0)}\,,
\end{eqnarray}
where the terms with $n<0$ are cut off by the highest weight property
of $U$ and the terms with $n>(p-1)h_\chi$ are cut off by the assumption 
that $\eta$ vanishes on $O_{(\infty^p)}^R$.  
This formula motivates the proof of:

\begin{lem} \label{inject} Let $\HH^R$ be any representation of the
conformal field theory that is generated from a highest weight state
$U$.  Then there is an injection
\begin{equation} \label{isom} 
\sigma: (A_{[p]}^R)^* \hookrightarrow (A_{[p,1]})^*\,.
\end{equation}
\end{lem}

\oproof{} We realize $A_{[p]}^R$ as $A_{(\infty^p)}^R$ and $A_{[p,1]}$
as $A_{(\infty^p,-1)}$.  Then define $\sigma$, as suggested above, by
the formula
\begin{equation}
\left[\sigma(\eta)\right] (\chi) = \eta(V(\chi,1)U) = 
\sum_{n = -(p-1)h_\chi}^0 \eta (V_n(\chi) U)\,. 
\end{equation}
In order to check that $\sigma(\eta)$ annihilates $O_{(\infty^p,-1)}$,
we observe from \eqref{defO} that $O_{(\infty^p,-1)}$ is generated by
the states of the form $V_{(\infty^p,-1)}^{(M)}(\psi)\chi$, where
$M>0$ and 
\begin{equation}\label{def1}
V_{(\infty^p,-1)}^{(M)}(\psi) = \oint_0
\frac{d\zeta}{\zeta^{M+1}} V\left[
\left(\frac{(\zeta+1)}{\zeta^{p-1}}\right)^{L_0} \psi,\zeta \right] \,.
\end{equation}
It is therefore sufficient to show that for $M>0$,
\begin{equation}\label{show1}
\eta\left(V(V_{(\infty^p,-1)}^{(M)}(\psi)\chi,1) U\right)=0 \,, 
\end{equation}
provided that $\eta\in (A_{(\infty^p)}^R)^\ast$. Expanding out
(\ref{def1}) in terms of modes we have
\begin{equation}
V(V_{(\infty^p,-1)}^{(M)}(\psi)\chi,1) =
\sum_{s=0}^{h_\psi} \binom{h_\psi}{s} 
V(V_{(-(p-1)h_\psi+s-M-1)}(\psi)\chi,1)
\,.
\end{equation}
Since the vertex operator is evaluated at $z=1$, we can rewrite it in
terms of a sum over all modes $V_{(r)}(\cdot)$. We then collect
together all those terms that have the same conformal weight: this
amounts to choosing $r$ (as a function of $s$) as  
$r=p h_\psi +  h_\chi + M -s + t$, 
where now $t$ labels the 
different values for the conformal weight of the resulting state. We
then apply Lemma~\ref{lem2} to 
$V_{(h_\chi+ph_\psi+M-s+t)}(V_{(-(p-1)h_\psi+s-M-1)}(\psi)\chi)$. The
first sum contains only terms of the form $V_{(-R)}(\psi)\phi$ with 
$R\geq (p-1)h_\psi-s+M+1$, for which $\eta$ vanishes by
assumption. The second sum gives rise to 
\begin{equation}
(-1)^{M+(p-1)h_\psi} \sum_{s=0}^{h_\psi} (-1)^{s} \binom{h_\psi}{s} 
\sum_{L\geq 0} \binom{(p-1)h_\psi + M -s + L}{L} 
V_{(h_\psi+h_\chi-1+t-L)}(\chi) V_{(L)}(\psi)\,.
\end{equation}
All terms with $L\geq h_\psi$ vanish since $V_{(L)}(\psi) U=0$ as $U$
is a highest weight state. It therefore only remains to check that all
the other terms vanish, \ie\ that 
\begin{equation}\label{claim1}
\sum_{s=0}^{h_\psi} (-1)^{s} \binom{h_\psi}{s} 
\binom{(p-1)h_\psi + M -s + L}{L} = 0 \quad 
\hbox{for $L=0,\dots,h_\psi-1$.} 
\end{equation}
In order to prove this identity, we observe that
\begin{equation}\label{consider}
\begin{split}
\sum_{s=0}^{h_\psi} (-1)^{s} & \binom{h_\psi}{s} 
\sum_{L\geq 0} \binom{(p-1)h_\psi + M -s + L}{L} u^L \\
& = \sum_{s=0}^{h_\psi} (-1)^{s} \binom{h_\psi}{s} 
\frac{1}{(1-u)^{(p-1)h_\psi+M-s+1}} \\
& = \frac{1}{(1-u)^{(p-1)h_\psi+M+1}}
\sum_{s=0}^{h_\psi} (-1)^{s} \binom{h_\psi}{s} (1-u)^s \\
& = \frac{u^{h_\psi}}{(1-u)^{(p-1)h_\psi+M+1}}\,.
\end{split}
\end{equation}
Thus the left-hand-side of (\ref{consider}) does not have any powers
of $u$ below $h_\psi$, and therefore (\ref{claim1}) holds.

To complete the proof we must check that $\sigma$ is injective, \ie\
that $\sigma(\eta) = 0$ implies $\eta=0$.  By Theorem~\ref{zthm}, 
$\sigma(\eta) = 0$ means that 
$\IP{\prod_{j} V(\psi_j,z_j)}_{\sigma(\eta)} = 0$ for all $\psi_j$ and
$z_j$; and since we can generate any mode acting on $U$ by taking
suitable contour integrals of vertex operators, it follows that $\eta$
annihilates any state generated from $U$.  But since $U$ generates the
whole of $\HH^R$ this implies that $\eta = 0$.  This completes the
proof. \done  

\noindent Combining Lemma~\ref{inject} and Theorem~\ref{findim} we now
obtain the desired result:

\begin{thm} \label{quasirat} Suppose $A_{[2]}$ is finite-dimensional.
Then every irreducible highest weight representation of the conformal
field theory is quasirational.
\end{thm}
\oproof{} Using Theorem~\ref{findim} and Lemma~\ref{ptsplit} we see
that $A_{[p,1]}$ is finite-dimensional for any $p\geq 1$.  Then from
Lemma~\ref{inject} it follows that each $A_{[p]}^R$ is
finite-dimensional, and the case $p=2$ implies that the special
subspaces are finite-dimensional. \done

Finally we observe that the tools we have developed here also allow
us to prove that the $C_2$ condition implies the finiteness of the
fusion rules: 

\begin{cor} \label{fusion} Suppose $A_{[2]}$ is finite-dimensional
and let $\HH^{R_i}$, $\HH^{R_j}$ and $\HH^{R_k}$ be three highest
weight representations of the conformal field theory. Then the fusion 
rule coefficient $N_{ij}^k$ is finite. 
\end{cor}
\oproof{} From the perspective of correlation functions what we are
claiming is that there are only finitely many ways to couple the three
highest weight representations; this follows from the
finite-dimensionality of $A_{[1,1,1]}^*$, and hence is a consequence
of Theorem~\ref{findim}. On the other hand, there are also more
algebraic approaches to fusion products \cite{FZ,Li2}; in lieu of
proving that these approaches are equivalent, we remark that it is
known \cite{FZ,Li3} that   
\begin{equation}
N_{ij}^k \le \dim \Hom_\A (A^{R_i}_{[1,1]} \otimes_\A A^{R_j}_{[1]}, 
A^{R_k}_{[1]}) \,.
\end{equation}
Since all spaces involved are finite-dimensional we get the desired 
result. \done

\section{A bound on the central charge} \label{centralcharge}

Up to now we have analysed what follows from the $C_2$ condition of
Zhu.  As we have seen, this assumption is already sufficient to prove
Nahm's conjecture.  If we assume in addition that $\A$ is
semisimple, then using Zhu's result about the modular properties of the
characters (see (ii) in Section~\ref{ratl}) we can derive a bound on
the effective central charge of the $W$-algebra. If $c$ denotes the
central charge of the Virasoro algebra, the effective central charge,
$\tilde{c}$, is defined to be $\tilde{c}=c-24 h_{min}$, where
$h_{min}$ is the smallest conformal weight of any state in any
(irreducible) highest weight representation of the theory. We can now
prove 

\begin{prop} \label{cbo} Suppose $A_{[2]}$ is finite-dimensional and
$\A$ is semisimple.  Then 
\begin{equation}\label{cbound}
\tilde{c} \leq \frac{(\dim A_{[2]} - 1)}{2} \,.
\end{equation}
\end{prop}

\oproof{} As in the proof of
Proposition~\ref{findimm}, let $k = \dim A_{[2]} - 1$, and define 
\begin{equation}
f_2(q) = \sqrt{2} q^{\frac{1}{24}} \prod_{n=1}^{\infty} (1+q^n)\,.
\end{equation}
This notation goes back to \cite{Pol}, although we have deviated
slightly from their convention by replacing $q^2$ with $q$; we could
also write $f_2$ in terms of conventional theta functions. In terms of
this function we can then rewrite \eqref{ineqq} as 
\begin{equation}\label{ineqqq}
\tr_{\HH}\;  q^{L_0} \le 2^{-\frac{k}{2}} q^{-\frac{k}{24}} f_2(q)^k \,.
\end{equation}
Here and in the following we shall always assume that $0<q<1$.

Next we follow closely an argument from \cite{EFHHNV}, using the
modular transformation properties of characters that were proven by
Zhu \cite{Zhu}. (As pointed out in \cite{KL}, that proof actually only
required the assumptions of the Proposition; by the way, this is the
only place where we use the semisimplicity of $\A$.)  If we write 
$q=e^{2\pi i\tau}$, and $\tilde{q}=e^{-2\pi i/\tau}$, then we have
\begin{equation} \label{modtra}
\chi_0(\tilde{q}) = \sum_{R} a_{R} \chi_{R}(q) \,,
\end{equation}
where the $a_{R}$ are some coefficients, $\chi_0$ is the character
of the vacuum representation, $\chi_R$ is the character of the
representation $\HH^R$, and the sum is over all irreducible
representations of Zhu's algebra. On the other hand, using the modular
transformation properties of $f_2$ (see for example \cite{Pol}) and
\eqref{ineqqq} we have 
\begin{eqnarray} \label{direct}
\chi_0(\tilde{q}) & \le & \tilde{q}^{-\frac{(k+c)}{24}}
                  2^{-\frac{k}{2}} f_2(\tilde{q})^k \\ 
& = & \tilde{q}^{-\frac{(k+c)}{24}} 2^{-\frac{k}{2}} f_4(q)^k\,,
\end{eqnarray}
where
\begin{equation}
f_4(q) = q^{-\frac{1}{48}} \prod_{n=1}^{\infty}
(1-q^{n-\frac{1}{2}})\,. 
\end{equation}
In the limit $\tau \to i\infty$ ($q \to 0, \tilde{q} \to 1$), 
\eqref{modtra} implies that 
\begin{equation} \label{c1}
\chi_0(\tilde{q}) = q^{h_{min} - \frac{c}{24}}(a + o(1))\,,
\end{equation}
where $a$ may be zero, while from \eqref{direct} we get
\begin{equation} \label{c2}
\chi_0(\tilde{q}) \le 2^{-\frac{k}{2}} (q^{-\frac{1}{48}} + O(q))^k 
= 2^{-\frac{k}{2}}q^{-\frac{k}{48}} (1+O(q))\,.
\end{equation}
Comparing \eqref{c1} and \eqref{c2} we get the desired result
$\tilde{c} \le k/2$.  \done

Incidentally, this proposition makes it clear that the dimension of
$A_{[2]}$ will often be bigger than that of $A_{[1,1]}$. For example,
for a self-dual theory we have $\dim A_{[1,1]}=1$, but the
proposition implies that $\dim A_{[2]} \geq 2\tilde{c}+1$. (For the
$e_8$ theory at level $1$, $\tilde{c}=8$, and thus we have that
$\dim A_{[2]} \geq 17$. As a matter of fact, we have checked that 
$\dim A_{[2]} \geq 4124$.)

To a physicist, the above argument can be explained as follows. Recall
that a basis for a theory of $k$ free R fermions is given by the
states 
\begin{equation}
\psi_{-N_1}^{i_1} \cdots \psi_{-N_n}^{i_n} \Omega\,,
\end{equation}
where $N_1 > \cdots > N_n > 0$. Comparing this with \eqref{spanset}
one might loosely say that the number of degrees of freedom of our
theory is bounded above by the number of degrees of freedom in a
theory of $k$ free fermions. The effective central charge measures in
essence the number of degrees of freedom; since every free fermion
contributes $1/2$, this explains the bound $\tilde{c} \le k/2$.

The original argument of \cite{EFHHNV} was very similar to that
presented above, except that they began with a spanning set
\eqref{spanset} where $N_1 \ge N_2 \ge \cdots \ge N_n$ and repeats
\ti{are} allowed. In essence, they were therefore comparing the theory
to a theory of $m$ free bosons (where $m$ is the dimension of the
generating set). The modular argument then involved the $\eta$
function (rather than the $f_2$ function), and the bound they obtained
was $\tilde{c}<m$. For theories for which an explicit (small) 
generating set is known, their bound tends to be stronger than
\eqref{cbound}, although not even this is the case in general: for the
$c=-22/5$ minimal model, our bound is $\tilde{c} \le 1/2$ while the
bound in \cite{EFHHNV} is $\tilde{c} < 1$; in actual fact 
$\tilde{c}= 2/5$ for this example. At any rate, Proposition~\ref{cbo} 
gives a bound on the effective central charge in terms of an intrinsic
quantity of the vertex operator algebra that can be easily
determined.

\section{An interpretation of $A_{[p,1]}$} \label{interp}

Finally we would like to give a more precise interpretation of the
spaces $A_{[p,1]}$: namely, we show that any correlation function of
the type described by $A_{[p,1]}^*$ is in fact obtained by inserting
one highest weight state and one state annihilated by all $V_n(\psi)$
with $n>(p-1)h_\psi$.  To prove this result, strengthening
Lemma~\ref{inject}, we will need to make a rather strong assumption on
the theory: namely, we assume that every weak module is completely
reducible into irreducible modules (this property has been called
\ti{regularity} in the literature on vertex operator algebras; in
particular, it was shown in \cite{Li} that regularity actually implies
$\dim A_{[2]} < \infty$.)  Then we can prove

\begin{prop} \label{dict}  Suppose every weak module is
completely reducible.  Then
\begin{equation}
\bigoplus_R (A_{[p]}^R)^* \otimes R \iso A_{[p,1]}^*\,,
\end{equation}
where the sum runs over all irreducible representations of Zhu's
algebra $\A$.
\end{prop}
\oproof{} We claim that the isomorphism is implemented by the
map
\begin{equation} \label{sdef}
\left[\sigma(\eta \otimes U)\right] (\chi) 
= \sum_{n = -(p-1)h_\chi}^0 \eta(V_n(\chi) U)\,.
\end{equation}
The calculation in the proof of Lemma~\ref{inject} demonstrates that
$\sigma$, as given in \eqref{sdef}, is well defined. In order to prove
that $\sigma$ is injective, we note that the argument in the proof of
Lemma~\ref{inject} shows that $\sigma(\eta \otimes U) = 0$ only if
$\eta \otimes U = 0$.  Now suppose $\sigma$ is not injective.  Then
there exists some linear dependence
\begin{equation} \label{ldep}
\sum_{i=1}^m \sigma(\eta_i \otimes U_i) = 0\,.
\end{equation}
Choose such a dependence with the smallest possible $m$; we have
already observed that $m = 1$ is impossible.  If $m>1$ then $U_1$ and
$U_2$ cannot be linearly dependent (else we could easily reduce $m$,
contradicting the minimality.) The complete reducibility implies that 
Zhu's algebra is semisimple \cite{Zhu}, and \eqref{wedd} then
guarantees that there exists some $a \in \A$ with $aU_1=0$, 
$aU_2\ne 0$; equivalently, there exists some $\psi \in \HH$ such that 
$V_0(\psi)U_1 = 0$, $V_0(\psi)U_2 \ne 0$.  Next we use
Theorem~\ref{zthm} to identify $A_{(\infty^p,-1)}^*$ with a space of
correlation functions. We can therefore re-express \eqref{ldep} as the
statement that
\begin{equation}
\sum_{i=1}^m \IP{\prod_{j} V(\psi_j,z_j)}_{\sigma(\eta_i \otimes U_i)}
= 0\,, 
\end{equation}
for all $\psi_j$ and $z_j$. By taking a suitable contour integral this
implies in particular that 
\begin{equation}
\sum_{i=1}^m 
\IP{\prod_{j} V(\psi_j,z_j) V_0(\psi)}_{\sigma(\eta_i \otimes U_i)} =
0\,, 
\end{equation}
and therefore that
\begin{equation}
\sum_{i=2}^m \sigma(\eta_i \otimes aU_i) = 0\,,
\end{equation}
contradicting the minimality of $m$.  This completes the proof of the 
injectivity.

It remains to show that $\sigma$ is surjective. Because of
Theorem~\ref{aacts} Zhu's algebra $\A$ acts on $A^*_{(\infty^p,-1)}$ via
its action at $-1$, and we can therefore decompose $A^*_{(\infty^p,-1)}$
as 
\begin{equation} \label{decomp}
A^*_{(\infty^p,-1)} = \bigoplus_{R} B_{[p]}^R \otimes R \,,
\end{equation}
where $B_{[p]}^R$ denotes an as yet undetermined multiplicity space.
Using Theorem~\ref{zthm} we can regard $A_{(\infty^p,-1)}^*$ as the
space of correlation functions $\IP{\prod_{j} V(\psi_j,z_j)}_\eta$,
satisfying the conditions

\begin{equation} \label{constr}
\begin{align}
\IP{\prod_{j} V(\psi_j,z_j) V_n(\psi)}_\eta & = 0  \for n > 0\,, \\
\IP{V_n(\psi) \prod_{j} V(\psi_j,z_j)}_\eta & = 0 
\for n < -(p-1) h_\psi\,.\label{constrone}
\end{align}
\end{equation}
Using the decomposition \eqref{decomp}, $B_{[p]}^R$ can
then be regarded as the space of correlation functions for which the
zero modes in (\ref{constr}) transform in the representation $R$ of
$\A$. Each element of $B_{[p]}^R$ defines a representation of the
conformal field theory where the state at $-1$ is a highest weight
state (that transforms in the representation $R$ under the action of
the zero modes), whereas the state at $\infty$ is only annihilated by
the modes $V_n(\psi)$ with $n>(p-1)h_\psi$. 

Now we would like to argue that each $\xi \in B_{[p]}^R$ actually
defines a linear functional on $\HH^R$, the Fock space generated by
the action of the modes on the highest weight state at $-1$. We might
\ti{a priori} worry that the correlation functions associated with
$\xi$ did not respect the null-vector relations by which one quotients
in the definition of $\HH^R$; indeed, in the definition of $\HH^R$ we
divided out states that vanish in amplitudes involving an arbitrary
number of vertex operators and a highest weight state in the (dual)
representation, but now we are considering what seem to be more
general amplitudes. To resolve this difficulty we use our extra
assumption of complete reducibility.  The condition \eqref{constrone}
is sufficient to deduce that the Fock space that is generated by the
action of the modes on the state at $\infty$ defines a weak
module, and therefore must be completely reducible into a direct sum
of irreducible highest weight representations. Thus in fact we are
only considering amplitudes where, apart from an arbitrary number of
vertex operators, we have a highest weight state at $\infty$, and 
therefore $\xi \in B_{[p]}^R$ indeed defines a linear functional
on $\HH^R$.  It follows from \eqref{constrone} that this functional
vanishes on $O^R_{(\infty^p)}$, and hence that it can be regarded as a
linear functional on  $A^R_{(\infty^p)} \iso A^R_{[p]}$.  It therefore
follows that $B_{[p]}^R\iso(A^R_{[p]})^*$, and we have thus
established the proposition. \done  

Proposition~\ref{dict} implies in particular that the dimension of
the quotient spaces $A^R_{[p]}$ for each representation $\HH^R$ is
bounded in terms of the quotient space $A_{[p+1]}$ of the vacuum
representation.  This result reflects the familiar fact that, for
rational theories, the vacuum representation already contains a
substantial amount of information about all representation spaces
$\HH^R$.

\section{Conclusions}\label{conclus}

In this paper we have proven the conjecture of Nahm that every
representation of a rational conformal field theory is
quasirational (Theorem~\ref{quasirat}). More specifically, we have
shown that if the conformal field theory satisfies the $C_2$ condition
of Zhu, \ie\ if the space $A_{[2]}$ is finite-dimensional, then the
quotient space $A^R_{[p]}$ of each highest weight representation
$\HH^R$ is finite-dimensional for $p\geq 1$; this immediately implies
that $\HH^R$ is quasirational. We have also shown that this implies
that the theory has only finitely many $n$-point functions, and in
particular that the fusion rules between irreducible representations
are finite (Corollary~\ref{fusion}). The main
technical result of the paper is the spanning set for the vacuum
representation of  any conformal field theory
(Proposition~\ref{core}), from which we have also been able to deduce 
various other properties of conformal field theories that 
satisfy the $C_2$ condition of Zhu (Corollary~\ref{finalg} and
Proposition~\ref{findimm}). 

We have introduced systematically spaces $A_{\bf u}$ that describe the
correlation functions with $k$ highest weight states at
$u_1,\dots,u_k$. Some of the structure of these spaces does not
depend on whether the $u_i$ are pairwise distinct, and one may
therefore hope that these spaces will be useful in extending the
definition of conformal field theory to singular limits, as envisaged 
in the program of Friedan \& Shenker \cite{FS}. 

In \cite{N} it was shown that the finite-dimensionality of $A_{[n]}$ 
implies the existence of $n$-point functions satisfying the
Knizhnik-Zamolodchikov equation.  Given Theorem~\ref{findim}, it
now follows that the existence of $n$-point functions already follows 
from the finite-dimensionality of $A_{[2]}$. Similarly, the condition
that $A_{(\infty^k)}$ is finite-dimensional in Theorem~\ref{ptindep}
can now be relaxed to the assumption that $A_{[2]}$ is
finite-dimensional. 

It may be possible to prove an inhomogeneous version of the finiteness
lemma (Proposition~\ref{core}). In particular, one may be able to
prove that the finite dimensionality of $\A$ implies the finite
dimensionality of all $A_{[1,1,\dots,1]}$. This would go a certain
way to proving (a version of) Zhu's conjecture, that the finite
dimensionality of Zhu's algebra implies that the $C_2$ condition is
satisfied. 

However, it seems likely that this will require more sophisticated
methods, since the conjecture apparently does not hold for meromorphic
field theories (that are not conformal). Consider the theory for which
$V$ is spanned by states $J^{a,i}$ of grade $1$, where
$a=1,\dots,248$ labels the adjoint representation of $e_8$,  
and $i\in\I$, where $\I$ is some countably infinite set.  For any
finite set of vectors in $V$ we can define the amplitudes to be the
products of the amplitudes that are associated to the different copies
of the affine $e_8$ theory at level $1$. These amplitudes are well
defined and satisfy all the conditions of \cite{GG} (except that the
theory does not have a conformal structure and the weight spaces are
not finite-dimensional). Since each $e_8$ level $1$ theory is
self-dual, it is easy to see that the same holds for the infinite
tensor theory; thus $\A$ is one-dimensional. However, the eigenspace
at conformal weight $1$ is infinite-dimensional, and
Proposition~\ref{findimm} therefore implies that the $C_2$ condition  
cannot be satisfied. On the other hand, most of our arguments (in
particular all of Section~\ref{basislem} and \ref{nconj}) do not
require a conformal structure or the assumption that the $L_0$
eigenspaces are finite-dimensional.

\section*{Acknowledgements}

We are indebted to Peter Goddard for many useful conversations,
explanations and encouragement. We also thank Terry Gannon for a
helpful discussion and a careful reading of a draft version of this
paper, and Haisheng Li for making us aware of his important work
on the subject. M.R.G. is grateful to the Royal Society for a University
Research Fellowship, and A.N. gratefully acknowledges financial
support from the British Marshall Scholarship and an NDSEG Graduate
Fellowship.

\pagebreak

\appendix

\section{The action of Zhu's algebra on $A_\uu$}

In this appendix we want to prove (\ref{eq1}) and (\ref{eq2}). Both
these statements follow from straightforward calculations. 

\subsection{Proof of (\ref{eq1})}

Without loss of generality we may assume that $\psi_i$, $i=1,2$ are
both vectors of definite conformal weight $h_i$. 
Using (\ref{Npoint}), we can then write 
the commutator $[V_\uu^{(0)}(\psi_1), V_\uu^{(L)}(\psi_2)]$ as 
\begin{eqnarray}\label{2}
&&\oint_0\left\{\oint_\zeta V(\psi_1,z)V(\psi_2,\zeta) 
\left(\frac{\prod_{j=2}^{k}(z-u_j)}{z^{k-2}} \right)^{h_1} 
\frac{dz}{z} \right\}
\left(\frac{\prod_{j=2}^{k} (\zeta-u_j)}{\zeta^{k-2}} \right)^{h_2}  
\frac{d\zeta}{\zeta^{L+1}}\nonumber\\
&&=\oint_0\left\{\oint_\zeta
V(V(\psi_1,z-\zeta)\psi_2,\zeta)
\left(\frac{\prod_{j=2}^{k} (z-u_j)}{z^{k-2}} \right)^{h_1} 
\frac{dz}{z} \right\} \nonumber\\
&& \qquad\qquad\qquad
\left(\frac{\prod_{j=2}^{k} (\zeta-u_j)}{\zeta^{k-2}} \right)^{h_2}  
\frac{d\zeta}{\zeta^{L+1}}\nonumber\\
&&=\sum\limits_{m=0}^{h_1+h_2-1}
\oint_0 V(V_{m+1-h_1}(\psi_1)\psi_2,\zeta)
\left(\frac{\prod_{j=2}^{k} (\zeta-u_j)}{\zeta^{k-2}} \right)^{h_2}
\frac{d\zeta}{\zeta^{L+1}} \nonumber\\
&& \qquad\qquad\qquad \left\{\oint_\zeta (z-\zeta)^{-m-1}
\left(\frac{\prod_{j=2}^{k} (z-u_j)}{z^{k-2}} \right)^{h_1} 
\frac{dz}{z} \right\}\,.
\end{eqnarray}
The integral in brackets is 
\begin{equation}\label{3}
\begin{split}
\frac{1}{m!} & \left. \frac{d^m}{dz^m} 
\left[ \left(\frac{\prod_{j=2}^{k} (z-u_j)}{z^{k-2}} \right)^{h_1} 
\frac{1}{z} \right] \right|_{z=\zeta} \\
& = \frac{1}{m!} \sum_{s=0}^{m} \binom{m}{s} (-1)^s 
\frac{1}{\zeta^{1+s}} \,
\frac{d^{m-s}}{dz^{m-s}} 
\left. \left(\frac{\prod_{j=2}^{k}(z-u_j)}{z^{k-2}} \right)^{h_1} 
\right|_{z=\zeta}\,,
\end{split}
\end{equation}
and the last derivative is of the form
\begin{equation}
\frac{d^{m-s}}{dz^{m-s}} 
\left. \left(\frac{\prod_{j=2}^{k} (z-u_j)}{z^{k-2}} \right)^{h_1} 
\right|_{z=\zeta}
= \left(\frac{\prod_{j=2}^{k} (\zeta-u_j)}{\zeta^{k-2}} 
\right)^{h_1-m+s}
\left[ \binom{h_1}{m-s} + O(\zeta^{-1}) \right] \,,
\end{equation}
where the last bracket consists of a finite sum of terms. 
Thus (\ref{3}) becomes
\begin{equation}
\begin{split}
\sum_{s=0}^{m} & C_s
\left(\frac{\prod_{j=2}^{k}(\zeta-u_j)}{\zeta^{k-2}} \right)^{h_1-m-1}
\left(\frac{\prod_{j=2}^{k}(\zeta-u_j)}{\zeta^{k-1}} \right)^{1+s}
\left[1+O(\zeta^{-1}) \right] \\
&= \sum_{s=0}^{m} \widehat{C}_s
\left(\frac{\prod_{j=2}^{k}(\zeta-u_j)}{\zeta^{k-2}} \right)^{h_1-m-1}
\left[1+O(\zeta^{-1}) \right] \,,
\end{split}
\end{equation}
where $C_s$ and $\widehat{C}_s$ are some constants. Putting this back
into (\ref{2}) and observing that the conformal weight of
$V_{m+1-h_1}(\psi_1)\psi_2$ is $h_1+h_2-m-1$, we obtain the statement.

\subsection{Proof of (\ref{eq2})}

We rewrite the left-hand-side of (\ref{eq2}) as 
\begin{equation}\label{e1}
\begin{split}
& = \oint \frac{d\zeta}{\zeta} 
\oint \frac{dw}{w^{L+1}} (w+1)^{h_\psi}
V \left[ \left( \frac{\prod_{j=2}^{k}(\zeta - u_j)}{\zeta^{k-2}}
\right)^{L_0} V(\psi,w) \chi,\zeta\right] \\
& = \oint \frac{d\zeta}{\zeta} \oint \frac{dw}{w^{L+1}}
(w+1)^{h_\psi} \left( 
\frac{\prod_{j=2}^{k}(\zeta-u_j)}{\zeta^{k-2}}\right)^{h_\psi+h_\chi} 
V \left[ 
V\left(\psi, w 
\frac{\prod_{j=2}^{k}(\zeta - u_j)}{\zeta^{k-2}}\right)\chi, 
\zeta\right] \\
& = \oint_0 \frac{d\zeta}{\zeta} \oint_\zeta 
\frac{dz}{(z-\zeta)^{L+1}}
\left( \frac{\prod_{j=2}^{k}(\zeta - u_j)}{\zeta^{k-2}}
\right)^{h_\chi+L}
\left(\frac{\prod_{j=2}^{k}(\zeta - u_j)}{\zeta^{k-2}} 
+(z-\zeta)\right)^{h_\psi} 
V(\psi,z) V(\chi,\zeta)\,,
\end{split}
\end{equation}
where in the first two lines the integrals are taken over the region
$|\zeta|>|w|$, and we have substituted, in the last line, 
$$
z = w \frac{\prod_{j=2}^{k}(\zeta - u_j)}{\zeta^{k-2}} + \zeta\,.
$$
Using the usual contour deformation trick, the last line of
(\ref{e1}) can be written as the difference of two contour integrals
\begin{equation}\label{e2}
\begin{split}
& = 
\oint \oint_{\abs{z}>\abs{\zeta}} d\zeta dz
\frac{1}{\zeta (z-\zeta)^{L+1}}
\left( \frac{\prod_{j=2}^{k}(\zeta - u_j)}{\zeta^{k-2}}
\right)^{h_\chi+L}
\left(\frac{\prod_{j=2}^{k}(\zeta - u_j)}{\zeta^{k-2}} +(z-\zeta)
     \right)^{h_\psi} 
V(\psi,z) V(\chi,\zeta) \\
& \qquad - 
\oint \oint_{\abs{z}>\abs{\zeta}} d\zeta dz
\frac{1}{\zeta (z-\zeta)^{L+1}} 
\left( \frac{\prod_{j=2}^{k}(\zeta - u_j)}{\zeta^{k-2}}
\right)^{h_\chi+L} \\
& \hspace{6cm}
\left(\frac{\prod_{j=2}^{k}(\zeta - u_j)}{\zeta^{k-2}} +(z-\zeta)
     \right)^{h_\psi} 
V(\chi,\zeta) V(\psi,z)\,.
\end{split}
\end{equation}
The two terms can now be considered separately. In the second term we
write 
$$
\frac{1}{(z-\zeta)^{L+1}} = (-1)^{L+1} \frac{1}{\zeta^{L+1}}
\sum_{M=0}^{\infty} \binom{L+M}{M} \left(\frac{z}{\zeta}\right)^M \,,
$$
and observe that 
$$
\left(\frac{\prod_{j=2}^{k}(\zeta - u_j)}{\zeta^{k-2}} +(z-\zeta)
     \right)^{h_\psi} = 
z^{h_\psi} \left(1 + \frac{c_1}{z} +
     \OO\left(\frac{z}{\zeta}\right)\right) \,.
$$
The second term therefore consists of terms of the form
$V^{(M)}_\uu(\chi)\hat\phi$ with $M>0$, and therefore can be dropped. In
reaching this conclusion we have used that if $\phi$ is in the Fock
space, only finitely many powers of $\frac{z}{\zeta}$ contribute.

In the first term we now write
$$
\frac{dz}{(z-\zeta)^{L+1}} = \frac{dz}{z^{L+1}} 
\sum_{M=0}^{\infty} \binom{L+M}{M} \left(\frac{\zeta}{z}\right)^M \,,
$$
and observe that 
\begin{equation}
\begin{split}
\left(\frac{\prod_{j=2}^{k}(\zeta - u_j)}{\zeta^{k-2}} +(z-\zeta)
     \right)^{h_\psi} & =
\left(\frac{\prod_{j=2}^{k}(z - u_j)}{z^{k-2}}\right)^{h_\psi}
\left(\frac{z^{k-2} (z-\zeta)}{\prod_{j=2}^{k}(z - u_j)}
+ \frac{\prod_{j=2}^{k} \frac{(\zeta - u_j)}{(z -
     u_j)}}{\left(\frac{\zeta}{z}\right)^{k-2}}\right)^{h_\psi}
     \\ 
& = \left(\frac{\prod_{j=2}^{k}(z - u_j)}{z^{k-2}}\right)^{h_\psi}
\left[1 + \OO\left(\frac{\zeta}{z}\right)\right]\,.
\end{split}
\end{equation}
Putting this back into (\ref{e2}) proves (\ref{eq2}). Again,
we have used here that if $\phi$ is in the Fock space, only finitely
many powers of $\frac{\zeta}{z}$ contribute.

\end{document}